\documentclass[12pt]{article}
\usepackage{amssymb,amsmath,epsfig}
\allowdisplaybreaks

\begin{document}
\title{\textbf{Construction of Charged Cylindrical Gravastar-like Structures}}
\author{Z. Yousaf
\thanks{zeeshan.math@pu.edu.pk}\\
Department of Mathematics, University of the Punjab,
\\Quaid-i-Azam Campus, Lahore-54590, Pakistan}
\date{}
\maketitle
\begin{abstract}
In this work, we extend the work of gravastars to analyze
the role of electromagnetic field in $f(R,T)$ gravity. We consider the irrotational cylindrically symmetric geometry and established the $f(R,T)$ field equations and conservation laws. After considering charged exterior geometry, the mathematical quantities for evaluating Israel junction conditions are being calculated. The mass of the gravastar-like cylindrical structure is calculated through the equations of motion at the hypersurface in the presence of an electromagnetic field. The behavior of electric charge on the length of the thin shell, energy content, and entropy of gravastar is being studied graphically. We concluded that charge has an important role in the length of the thin shell, energy content, and entropy of gravastar.
\end{abstract}


\section{Introduction}

Gravitation is perhaps the basic interaction one can experience most
effectively in daily activities. General relativity, along
with the quantum theory, could be seen as the rise of modern
physics. The consequences of relativity must be brought into
consideration when examining galactic weak gravitational force.
There are eloquent relativistic sources, such as white dwarfs,
neutron stars, and black holes, in which these effects could have
significant results. As a matter of interest, it becomes absolutely
essential to follow observationally workable theories of gravity in
order to fully understand such systems. In addition, there are many
surprising results from the observations of type-Ia supernovae,
anisotropy data of our cosmos, etc.
\cite{1,2,3}
that have started a major revolution in gravitation,
thereby starting a new research platform. These disclosed that our
cosmos currently experiences an accelerating expansion.

A very reasonable approach among the theoretical physicists is the
implementation of modified gravitational theories (MGT) after
generalizing the Einstein-Hilbert (EH) action to investigate the dark secret of cosmological evolution. It is worthy to mention that
MGTs include $f(G)$ theory, where $G$ is a Gauss-Bonnet term
\cite{4}, $f(G,T)$
\cite{5,6,7}
(where $T$ is the trace of energy-momentum tensor), $f(R,\Box R, T)$ \cite{8,9} (where $\Box$ is the de Alembert's operator), and
$f(R,T,R_{\gamma\delta}T^{\gamma\delta})$
\cite{10,11,12,13,14},
etc. Several articles and reports exist in the literature on the
discussion of dark energy \cite{15,16}
and MGTs \cite{17,18,19,20,21,22,23,24,25}
as an elaboration for cosmic acceleration. The simplest alternative
version of GR is $f(R)$ theory, attained by displacing the Ricci
scalar $R$ in the EH action with its generic function. As a
a plausible explanation of cosmic inflation, this theory has been
source of great attraction among many researchers due to the
inclusion of quadratic $R$ terms \cite{26} in the EH action. Harko \emph{et al.} \cite{27} laid out the grounds for $f(R)$ gravity and the concept of $f(R,T)$ ($T$ is introduced due to quantum effects or exotic imperfect content), is introduced in which non-minimal coupling between geometry and matter was established. They formulated the corresponding equations of motion describing certain astrophysical perspectives.

Houndjo \cite{28} attempted to reconstruct
cosmological models in $f(R, T)$ theory and argued that his models
might feasibly explain accelerated and matter-dominated cosmic
histories. After employing the perturbation method on the mathematical
power-law models, Baffou \emph{et al.} \cite{29}
studied cosmic instability problems and determined their viability
requirements. Sahoo \emph{et al.} \cite{30} analyzed
phases of spatially homogeneous cosmos in $f(R,T)$ gravity.
Recently, Moraes \emph{et al.} \cite{31} inspected
the presence of stellar bodies in our accelerating universe and
managed to find some viability constraints in $R + \chi T$ gravity,
where $\chi$ is a constant. Yousaf \emph{et al.}
\cite{32} examined the role of $f(R,T)$ gravity after
subjecting the initially homogeneous compact structure into the
inhomogeneous one and found a relatively slow rate of collapse due to
$f(R,T)$ correction. They also provided relations connecting the
histories of inflationary and late-time acceleration of the
universe. Bhatti \emph{et al.}
\cite{33,34,35,36}
also worked in $f(R,T)$ cosmology for different astrophysical
models.

Gravitational collapse is amongst the most remarkable phenomena of the stellar dynamics.
The endpoint of gravitational collapse may result in black holes, neutron stars or white dwarfs. Although there
exists significant works that indicate that the collapse of
theoretically, acceptable matter contents may result in the formation
of naked singularity (NS). Herrera \emph{et al.}
\cite{37} examined NS formation in the spherical
relativistic matter distribution with the help of Weyl tensor.
Virbhadra \emph{et al.}
\cite{38,39} provided a
theoretical framework due to which one can study significantly
relations between the formation of NS and black holes as a result of
celestial collapse.

Mazur and Mottola \cite{40} proposed the
alternative concept of endpoints in the gravitational collapse of supermassive stars. They modified the notion of Bose-Einstein
for self-gravitating sources. They presented the gravitational vacuum star (gravastar), which is alternative to the black hole. Gravastar has no singularity and event horizon. It is comprised of three specific layers interior region, middle thin shell and exterior region. The middle shell is very thin with the finite range $r_1<r<r_2$. Here $r_1=r$ is the radius of interior and $r_2=r+\epsilon$ is the radius of the exterior region of gravastar, $\epsilon$ indicates the thickness of the middle shell, which is very very small. According to the Mazur and Mottola model, these layers of gravastar has three specific EoS. $\rho+p=0$ is the EoS for interior region, $\rho-p=0$ for thin shell and $\rho=p=0$ for
the exterior region.

Rahaman \emph{et al.} \cite{41} discussed charge region of gravastar under
(2+1)-dimensional space-time. They analyzed various theoretical feasible features,
like length, entropy, energy for the spherical configuration. Pani \emph{et al.} \cite{42} considered irrotational
thin-shell gravastar models by taking Schwarzschild and de Sitter
metric as exterior and interior geometries, respectively. They
performed linearized stability analysis in order to view the
existence of such structures in the cosmos. Cecilia \emph{et al.}
\cite{43} presented a detailed analysis in order
to study gravastars on the dynamical properties of fast rotating
those compact objects that do not possess a horizon. They declared
that not all gravastars which rotate are unstable and the origin of
gravastars is still highly speculative.

Cattoen \emph{et al.} \cite{44} presented the
usual concept of a gravastar and indicated that these could be
treated as an alternative to the black hole. They considered a family
of mathematical models characterizing with continuous pressure. They calculated the role of some physical variables, like anisotropy pressure, minimization of the anisotropic region, etc in the formation of gravastars at the boundary surface. Cardoso \emph{et al.} \cite{45} considered the properties of rotating gravastars with the help of the perturbation scheme and estimated their stability within the slow-rotation approximation. They examined the instability effects on those objects which do not have an event horizon like Kerr black hole, gravastars, and boson stars. They concluded that when these ultra-compact objects undergo rapid spinning, then a strong
ergoregion instability may appear in their dynamics.

Ghosh \emph{et al.} \cite{46} examined the possibility
of the existence of gravastars in higher dimensional spacetime and
inferred that few important indications about the formation of such
celestial structure with or without electric charge. Das \emph{et
al.} \cite{47} examined a family of singularity-free
gravastar models in $f(R,T)$ gravity and claimed that $f(R,T)$ gravity
could likely to host such structures. Shamir and Mustaq \cite{48} extended these results and examine the
role of $f(G,T)$ gravity on the possible formation of gravastars.
Recently, Yousaf \emph{et al.} \cite{49} presented
non-singular spherically symmetric gravastar model after evaluating
viability conditions in the presence of Maxwell-$f(R,T)$
corrections. Bhatti \cite{50} described the modeling of cylindrical gravastars-like structures in the field of GR.

This paper aims to analyze the role of Maxwell-$f(R,T)$
corrections on the possible modeling of gravastars-like cylindrical
structures, a viable alternative to the final stage of gravitational
collapse. The paper consists of the following sections. In the coming section, we briefly elaborate the basic framework of $f(R,T)$ theory and established the field equation of modified theory. Section \textbf{3} is devoted to computing modified field equations and mass of the gravastar with an EoS. In section \textbf{4}, we shall examine the effects of electromagnetic field on the viable matching conditions between exterior and interior geometries. In section \textbf{5}, we determine a few characteristics in cylindrical gravastar-like structures. In section \textbf{5}, we will discuss the role of electric charge on some physical valid properties of gravastar.

\section{$f(R,T)$ Theory and Charged Cylindrical System}

One can modify the EH action for $f(R,T)$ gravity as under
\begin{eqnarray}\label{1}
\textbf{S}=\frac{1}{16\pi}\int{f(R,\textit{T})}\sqrt{-g}~d^4x+\int\textit{\L}_m\sqrt{-g}~d^4x,
\end{eqnarray}
where $\textit{\L}_m$ describes the Lagrangian density for an ordinary
fluid, $g$ and $T$ are the traces of metric and the energy momentum
tensor, respectively. The energy momentum tensor through
$\textit{\L}_m$ can be defined as follows
\begin{eqnarray}\label{2}
T_{\zeta\eta}=\frac{-2}{\sqrt{-g}}\frac{\delta(\sqrt{-g}\textit{\L}_m)}{\delta
g^{\zeta\eta}}=g_{\zeta\eta}\textit{\L}_m-2\frac{\partial\textit{\L}_m}{\delta
g^{\zeta\eta}}.
\end{eqnarray}
Varying the action of $f(R,T)$ theory with respect to
metric tensor provides
\begin{eqnarray}\label{3}
\delta S=\frac{1}{16\pi}\int\left[f_R\delta R+f_T\frac{\delta
T}{\delta g^{\zeta\eta}}\delta
g^{\zeta\eta}-\frac{1}{2}g_{\zeta\eta}f\delta
g^{\zeta\eta}+16\pi\frac{1}{\sqrt{-g}}\frac{\delta(\sqrt{-g}\textit{\L}_m)}{\delta
g^{\zeta\eta}} \right]\sqrt{-g}~d^4x,
\end{eqnarray}
where $f_R(R,T)$ and $f_T(R,T)$ indicate the partial derivative with
respect to Ricci scalar and trace of energy momentum tensor
respectively. In Eq.\eqref{3}, one can find the value of $\delta R$
as under
\begin{eqnarray}\label{4x}
\delta R=\delta(g^{\zeta\eta}R_{\zeta\eta})=R_{\zeta\eta}\delta
g^{\zeta\eta}+g^{\zeta\eta}(\nabla_\gamma\delta\Gamma^\gamma_{\zeta\eta}-\nabla_\eta\delta\Gamma^\gamma_{\zeta\gamma}),
\end{eqnarray}
in which
\begin{eqnarray}\label{4ax}
\delta\Gamma^\gamma_{\zeta\eta}=\frac{1}{2}g^{\gamma\beta}(\nabla_\zeta
\delta g_{\eta\beta}+\nabla_\eta \delta g_{\beta\zeta}-\nabla_\beta
\delta g_{\zeta\eta}).
\end{eqnarray}
Using Eqs.\eqref{4ax} in Eq.\eqref{4x}, we get
\begin{eqnarray}\label{5}
\delta R=R_{\zeta\eta}\delta g^{\zeta\eta}+g_{\zeta\eta}\Box\delta
g^{\zeta\eta}-\nabla_\zeta\nabla_\eta\delta g^{\zeta\eta},
\end{eqnarray}
where $\Box\equiv \nabla_\alpha\nabla^\alpha$. Equation \eqref{5}
eventually makes Eq.\eqref{3} as under
\begin{eqnarray}\nonumber
\delta S&=&\frac{1}{16\pi}\int\left[f_R(R,T)R_{\zeta\eta}\delta
g^{\zeta\eta}+f_R(R,T)g_{\zeta\eta}\Box\delta
g^{\zeta\eta}-f_R(R,T)\nabla_\zeta\nabla_\eta\delta
g^{\zeta\eta}\right.\\\label{6}&+&\left.f_T(R,T)\frac{\delta(g^{\mu\nu}T_{\mu\nu})}{\delta
g^{\zeta\eta}}\delta
g^{\zeta\eta}-\frac{1}{2}g_{\zeta\eta}f(R,T)\delta
g^{\zeta\eta}+\frac{16\pi}{\sqrt{-g}}\frac{\delta(\sqrt{-g}\textit{\L}_m)}{\delta
g^{\zeta\eta}}\right]\sqrt{-g}d^4x.
\end{eqnarray}
After some manipulations, one can write the corresponding equation
of motion as follows
\begin{eqnarray}\label{7}
f_{R}(R_{\zeta\eta}-\nabla_\zeta\nabla_\eta+g_{\zeta\eta}\Box)+f_{T}(T_{\zeta\eta}+\Theta_{\zeta\eta})=8\pi
T_{\zeta\eta}+\frac{1}{2}f g_{\zeta\eta},
\end{eqnarray}
where
\begin{eqnarray}\label{8}
\Theta_{\zeta\eta}=\frac{g^{\mu\upsilon}\partial
T_{\mu\upsilon}}{\partial g^{\zeta\eta}}.
\end{eqnarray}

An important point that must be stressed here is that the divergence
of an energy-momentum tensor in this theory, unlike GR and $f(R)$,
is non-zero. It can be given as follows
\begin{eqnarray}\label{9}
\nabla^\zeta
T_{\zeta\eta}=\frac{f_{T}(R,\textit{T})}{8\pi-f_{T}(R,\textit{T})}[(T_{\zeta\eta}
+\Theta_{\zeta\eta})\nabla^\zeta\ln{f_T(R,\textit{T})}+\nabla^\zeta(\Theta_{\zeta\eta})].
\end{eqnarray}
which allows this theory to break both weak and strong equivalence
principles. Consequently, one can observe the non-geodesic motion of
the test particles. It is possible to recover $f(R)$ gravity under
the constraint $f(T)=0$.

The aim of the present work is to analyze the formation of
cylindrical gravastar-like objects in $f(R,T)$ gravity under the
influence of electromagnetic field. For this purpose we assume the
following static form of cylindrical metric
\begin{eqnarray}\label{13}
ds^2=-H(r)dt^2+K(r)dr^2+r^2(d\phi^2+\alpha^2dz^2),
\end{eqnarray}
where $H(r)=\sqrt{\alpha^2r^2-\frac{4M}{\alpha
r}+\frac{4q^2}{\alpha^2 r^2}}$ and $H(r)=\frac{1}{K(r)}$ with $M$
indicates the gravitating mass of the object while $q$ is the charge
and $\alpha$ is a constant term with the dimensions of inverse
length. We then model our source for the above geometry as the
locally-isotropic fluid configuration, whose mathematical form can
be expressed as under
\begin{eqnarray}\label{10}
T_{\zeta\eta}=(\rho+p) V_{\zeta}V_{\eta}+pg_{\zeta\eta},
\end{eqnarray}
where $V_{\zeta}$ is the four velocity, $\rho$ and $p$ denote the
density and pressure of the relativistic fluid, respectively, Under
comoving coordinate system, the vector $V^\mu$ satisfies
$V_{\zeta}V^{\zeta}=1$ relation. To solve the non-linear nature of
field equations, as described in
\cite{27,51}, we take
$\textit{\L}_m=-p+\mathcal{F}$ for the charged perfect fluid, where
$\mathcal{F}=-F_{\mu\nu}F_{\alpha\beta}g^{\mu\nu}g^{\alpha\beta}$=$-\frac{2q^2}{r^4}$,
for which $\Theta_{\zeta\eta}$ is found to be
$\Theta_{\zeta\eta}$=$-(2T_{\zeta\eta}+p
g_{\zeta\eta}+\mathcal{F}g_{\zeta\eta})$. Here $F_{\alpha\beta}$
stands for Maxwell tensor. To give mathematical as well as
cosmologically observational consistency to $f(R,T)$ theory, the
choice of $f(R,T)$ model is of key importance. Therefore, we
consider here $f(R,T)$=$R+2\chi T$, where $\chi$ is constant. This
simplest case of $f(R,T)$ could be fully equivalent to GR, after
rescaling the value of $\chi $ in the analysis. In this framework,
Eq.\eqref{7}, turns out to be
\begin{eqnarray}\label{11}
G_{\zeta\eta}=8\pi (T_{\zeta\eta}+E_{\zeta\eta})+\chi
T_{\zeta\eta}g_{\zeta\eta}+2\chi[T_{\zeta\eta}+E_{\zeta\eta}+pg_{\zeta\eta}+\mathcal{F}g_{\zeta\eta}],
\end{eqnarray}
where $G_{\zeta\eta}$ is an Einstein tensor and $E_{\zeta\eta}$
describes an energy momentum tensor for a charged matter. It can be
given as follows
\begin{equation}\label{4}
E_{\alpha\beta}=\frac{1}{4\pi}\left(F^{\gamma}_{\alpha}F_{\beta\gamma}
-\frac{1}{4}F^{\gamma\delta}F_{\gamma\delta}g_{\alpha\beta}\right).
\end{equation}
In this background, Eq.\eqref{9} takes the following form
\begin{eqnarray}\label{12}
\nabla^\zeta
T_{\zeta\eta}=\frac{-\chi}{2(4\pi+\chi)}\left[2\nabla^\zeta(pg_{\zeta\eta})+2\nabla^\zeta(\mathcal{F}g_{\zeta\eta})+g_{\zeta\eta}\nabla^\zeta
T\right].
\end{eqnarray}
By the substitution of $\chi=0$ in the above equation, one can
simply obtain the GR results.

The non-zero components of Einstein tensor for the cylindrically
symmetric interior geometry are
\begin{align}\label{14}
G^0_0&=-\frac{1}{K^2 r^2}(\acute{K}r-K),\\\label{15}
G^1_1&=\frac{1}{KHr^2}(\acute{H}r+H),\\\label{16}
G^2_2&=\frac{1}{4H^2K^2r}(2\acute{\acute{H}}HK-H\acute{H}\acute{K}-K\acute{H}^2r-2\acute{K}H^2+2\acute{H}KH).
\end{align}
Making use of Eqs.\eqref{13}-\eqref{16} in Eq.\eqref{11}, we obtain
the following equations
\begin{align}\label{17}
&\frac{\acute{K}r-K}{K^2}=-r^2[8\pi\rho-\chi(p-3\rho)]-\frac{q^2}{r^2}\left[1+\chi\left(\frac{1}{4\pi}-4\right)\right],\\\label{18}
&\frac{\acute{H}r+H}{H K}= -r^2[8\pi
p+\chi(3p-\rho)]+\frac{q^2}{r^2}\left[1+\chi\left(\frac{1}{4\pi}-4\right)\right],\\\nonumber
&\frac{r}{4H^2K^2}(2H'')H
K-H\acute{H}\acute{K}-K\acute{H}^2r-2\acute{K}H^2+2\acute{H}K
H)\\\label{19} &=-r^2[8\pi
p+\chi(3p-\rho)]+\frac{q^2}{r^2}\left[1+\chi\left(\frac{1}{4\pi}-4\right)\right].
\end{align}
Here, we noticed that $G_{33}=\alpha^2G_{22}$. For $f(R,T)$ theory,
the hydrostatic equilibrium condition satisfy Eq.\eqref{9} and is
defined as
\begin{eqnarray}\label{20}
\frac{\acute{H}}{2H}(\rho+p)+\frac{d p}{dr}-\frac{q^2}{2\pi
r^5}-\frac{\chi}{(4\pi+\chi)}\left[\frac{1}{2}(\acute{p}-\acute{\rho})-\frac{8q^2}{r^5}\right]=0.
\end{eqnarray}
Using Eq.\eqref{17}, we get
\begin{eqnarray}\label{21}
\frac{1}{K}=\frac{8m}{3h}+\chi(\rho-\frac{p}{3})r^2+\frac{q^2}{r^2}\left[1+\chi\left(\frac{1}{4\pi}-4\right)\right],
\end{eqnarray}
where $r$ represents the radius of interior region of gravastar and
$m$ is the corresponding gravitational mass within the interior
region. Substituting Eq.\eqref{21} in Eq.\eqref{20}, one can obtain
\begin{eqnarray}\label{22}
\frac{d p}{dr}=\frac{\frac{q^2}{2\pi
r^5}-\frac{8q^2\chi}{r^5(4\pi+\chi)}-\frac{H'}{2H}(\rho+p)}{[1+\frac{\chi}{2(4\pi+\chi)}(1-\frac{d\rho}{d
p})]},
\end{eqnarray}
where
\begin{eqnarray*}
\frac{H'}{H}=K r[-8\pi p+\chi(\rho-3 p)]+\frac{K
q^2}{r^3}\left[1+\chi\left(\frac{1}{4\pi}-4\right)\right]-\frac{1}{r}.
\end{eqnarray*}

\section{Structure of a Gravastar}

It is well established that the exterior and interior geometries of gravastars are supported by particular forms of EoS. We
consider that geometric structure of an interior exterior spacetime
is described by $p=-\rho$ relation. One can notice that this EoS
describes the contribution of dark energy in the modeling
of the interior region. This is a particular case of barotropic EoS
$p=\omega \rho$. Equation \eqref{22} could be used to prove the
following relation
\begin{eqnarray}\label{23}
p=-\rho_0(constant),
\end{eqnarray}
Using Eq.\eqref{23} in Eq.\eqref{17}, we get
\begin{eqnarray}\label{24}
\frac{1}{K}=\frac{r^2}{3}[4(2\pi+\chi)Y_0]-\frac{q^2}{r^2}\left[1+\chi\left(\frac{1}{4\pi}-4\right)\right]
,~~~~ and~~~~K=W H^{-1},
\end{eqnarray}
where $W$ is an integration constant. The second of above equation describes a particular
relationship between metric coefficients of our observed spacetime. The
gravitational mass $M(D)$ for the interior region can be described as
\begin{eqnarray}\label{25}
M(D)=\int^{r_1=D}_04\pi
r^2\left(\rho_0+\frac{q^2}{2r^2}\right)dr=2\pi
D\left(\frac{2}{3}D^2+q^2\right).
\end{eqnarray}
This equation indicates that the gravitating matter content of the stellar body and its\ radial distance have a direct relationship, which can be supposed to be a very important feature of compact relativistic objects. This also shows the main dependence of $M$ on a particular radial coordinate. The above integral occupies the properties of improper integral on setting $r=\infty$. But the selection of this
radius $r$ is not realistic.

Now, we elaborate the geometry of middle thin shell and
discuss the effect of electromagnetic charge on the pressure within gravastar. Middle thin shell of gravastar is non-vacuum so it contain have a zero pressure. This zone of the gravastar is formed by the ultrarelativistic fluid. The EoS of the thin shell is $p=\rho$. At zero pressure, the solution of field equations is much difficult. To overcome this query, we used some approximation for finding an analytical solution, i.e., $0<H^{-1}\ll1$. Physically, we can say that when the exterior vacuum metric joins with the interior vacuum metric, they form a thin shell \cite{52}. Using EoS $\rho-p=0$ in Eqs.\eqref{17} and \eqref{18}, we obtain
\begin{eqnarray}\label{26}
\frac{d}{dr}(\ln
K)=-\frac{2q^2}{r^3}\left[1+\chi\left(\frac{1}{4\pi}-4\right)\right],
\end{eqnarray}
On substituting the EoS $\rho-p=0$ in Eqs.\eqref{17} and \eqref{19},
we get
\begin{eqnarray}\label{27}
-\frac{r}{2}\left(\frac{r\acute{H}}{2H}+1\right)\acute{K}=K^2,
\end{eqnarray}
which upon integrating Eq.\eqref{26} yields
\begin{eqnarray}\label{28}
\ln
K=\frac{q^2}{r^2}\left[1+\chi\left(\frac{1}{4\pi}-4\right)\right]+C.
\end{eqnarray}
\begin{eqnarray}\label{29}
\frac{\acute{H}}{H}=\frac{e^{\frac{q^2}{r^2}\left[1+\chi\left(\frac{1}{4\pi}
-4\right)\right]}}{\frac{2q^2}{r^3}\left[1+\chi\left(\frac{1}{4\pi}-4\right)\right]}-\frac{r}{2}.
\end{eqnarray}
Using EoS $p=\rho$ in Eq.\eqref{20}, we get
\begin{eqnarray}\label{30}
\frac{d}{dr}(\ln
p)=\frac{e^{\frac{q^2}{r^2}\left[1+\chi\left(\frac{1}{4\pi}-4\right)\right]}}
{\frac{2q^2}{r^3}\left[1+\chi\left(\frac{1}{4\pi}-4\right)\right]}-\frac{r}{2}.
\end{eqnarray}
Integration of the above equation provides
\begin{eqnarray}\label{30}
\rho=p=Exp\left[-\frac{r^4+Qr^2-Q}{8Q}e^{\frac{Q}{r^2}}+\frac{r^2}{4}\right],
\end{eqnarray}
where
$Q=q^2\left[1+\chi\left(\frac{1}{4\pi}-4\right)\right]$. One can note down that here density has a direct proportionality with the charge and radius. This shows that the ultrarelativistic fluid within the central shell of the gravastar is relativity denser with huge charge at the outer regions than that found in the inner regions of the stellar body. Further, when we move out from central point of the region, the more dense with high charged fluid can be noticed. The exterior region should obey EoS, $p=\rho=0$ and can be defined through an exterior charged spacetime, for details pleasee see \cite{50}.
\begin{center}\begin{figure}\centering
\epsfig{file=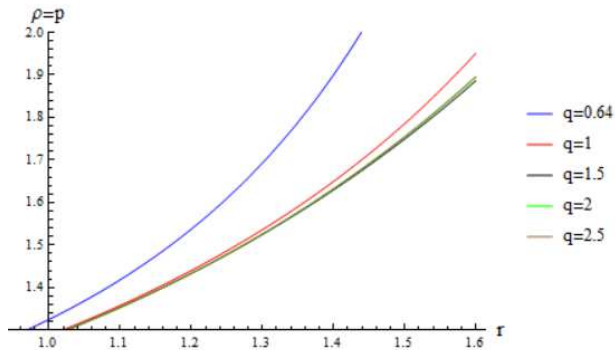,width=0.5\linewidth}\caption{\small{Plot
of $p=\rho$(km) verses $r$ in the gravastars with
$\chi=1$.}}\label{b1}
\end{figure}\end{center}

\section{Junction Condition between Interior and Exterior Geometries}

In this section, we formulate the junction condition by matching
interior and exterior regions and establish some relations that could allow the corresponding geometries to match smoothly on the hypersurface. It is worth notable that the formulation of gravastar is based on three
regions. Let $r_1$ and $r_2$ is the radii of interior and exterior
regions.  We will use Darmois and Israel \cite{52,53} junction conditions to formulate the smooth matching between interior and exterior surfaces. We will find out surface stress-energy tensor $S_m^n$ with the help of Darmois and Israel formulation. Lanczos equation is defined as \cite{54,55,56,57}
\begin{eqnarray}\label{32}
S_m^n=-\frac{1}{8\pi}(\ell_m^n-\delta_m^n \ell_i^i),
\end{eqnarray}
where $\ell_{mn}=L_{mn}^+-L_{mn}^-$, (+) and (-) signs indicate the
interior and exterior surfaces.\\
The second fundamental \cite{49,58,59,60} form (extrinsic curvature) is
defined by
\begin{align}\label{33}
\Omega_{\alpha\beta}^{\pm}&=\frac{\partial
x^\mu}{\partial\xi^\alpha}\frac{\partial
x^\nu}{\partial\xi^\beta}\nabla_\mu n_\nu, \\\label{34}
&=-n_\omega^{\pm}\left[\frac{\partial^2x_\omega}{\partial\xi^\alpha\partial\xi^\beta}+\Gamma_{\mu\nu}^\omega\frac{\partial
x^\mu}{\partial\xi^\alpha}\frac{\partial
x^\nu}{\partial\xi^\beta}\right]_\Sigma,
\end{align}
where $\xi^\alpha$ and $x^\mu$ can be determined through the
coordinates of hypersurface $(\Sigma)$ and manifold. Further, the vector
$n_\omega^{\pm}$ corresponds to the double-sided normals on the surface of
cylinder which can be defined as
\begin{eqnarray}\label{35}
n_\omega^{\pm}=\pm\frac{1}{\left|g^{\gamma\delta}\frac{\partial
g(r)}{\partial x^\gamma}\frac{\partial g(r)}{\partial
x^\delta}\right|^{\frac{1}{2}}}\frac{\partial g(r)}{\partial
x^\omega},
\end{eqnarray}
with timelike condition $n_\xi n^\xi=1$. The values of the energy-momentum tensor at the boundary surface turn out to be
$S_m^n=diag[\varrho,-\vartheta,-\vartheta,-\vartheta]$. Here
$\varrho$ is the surface density and $\vartheta$ is the pressure on
the surface. The expressions of $\vartheta$ and $\varrho$ are given
below
\begin{align}\label{36}
\varrho&=-\frac{1}{4\pi D}[\sqrt{g(r)}]_-^+,\\\label{37}
\vartheta&=-\frac{\vartheta}{2}+\frac{1}{16\pi}\left[\frac{g(r)'}{\sqrt{g(r)}}\right]_-^+.
\end{align}
Using Eqs.\eqref{36} and \eqref{37}, we get
\begin{eqnarray}\label{38}
\varrho=-\frac{1}{4\pi
D}\left[\sqrt{D^2\alpha^2-\frac{4M}{D\alpha}+\frac{4q^2}{D^2\alpha^2}}-\sqrt{\frac{4D^2}{3}(2\pi+\chi)\rho_0-\frac{Q}{r^2}}~
\right],
\end{eqnarray}
and
\begin{eqnarray}\label{39}
\vartheta=\frac{1}{8\pi
D}\left[\frac{3D^2\alpha^2-\frac{4q^2}{D^2\alpha^2}}{\sqrt{D^2\alpha^2-\frac{4M}{D\alpha}+\frac{4q^2}{D^2\alpha^2}}}
-\frac{3\rho_0D^2(2\pi+\chi)+\frac{Q}{r^2}-\frac{2Q}{r^3}}{\sqrt{\frac{4D^2}{3}(2\pi+\chi)\rho_0-\frac{Q}{r^2}}}\right].
\end{eqnarray}
Using the areal density, we can determine the mass of the thin shell as
\begin{eqnarray}\label{40}
m_s=4\pi
D^2\varrho=D\left[\sqrt{D^2\alpha^2-\frac{4M}{D\alpha}+\frac{4q^2}{D^2\alpha^2}}-\sqrt{\frac{4D^2}{3}(2\pi+\chi)\rho_0-\frac{Q}{r^2}}~
\right],
\end{eqnarray}
where
\begin{eqnarray}\label{41}
M=-\frac{m_s^2\alpha}{4D}-\frac{D^3\alpha}{3}(2\pi+\chi)\rho_0+\frac{D\alpha
Q}{4r^2}+\frac{D^3\alpha^2}{4}+\frac{q^2}{D\alpha}+\frac{m_s\alpha}{2}\sqrt{\frac{4D^2}{3}(2\pi+\chi)\rho_0-\frac{Q}{r^2}}.
\end{eqnarray}
is the total amount of quantity of matter within the charged gravastar.

\section{Few Characteristics of Cylindrical Gravastar-like Model}

In this section, we observe the effect of electromagnetic field on
some physical valid characteristic of gravastar.

\subsection{Proper Length of the Shell}

\begin{center}\begin{figure}\centering
\epsfig{file=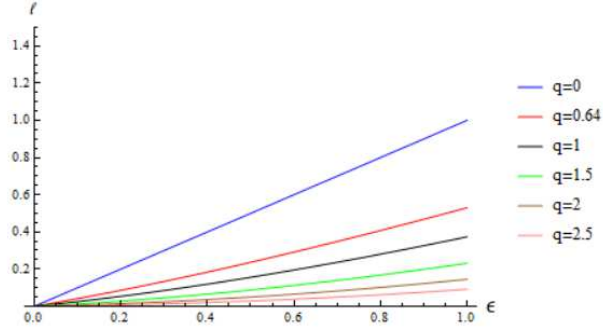,width=0.5\linewidth}\caption{\small{Plot
of $\ell$(km) verses $\epsilon$ in the gravastars with
$C=0.00006$.}}\label{b2}
\end{figure}\end{center}

Let us consider that $r_1=D$ is the radius of interior and
$r_2=D+\epsilon$ is the radius of exterior region of gravastar.
According to Mazur and Mottola \cite{40}, proper length of the
middle shell can be determined as
\begin{eqnarray}\label{42}
\ell=\int^{D+\epsilon}_D\sqrt{K}~dr=
\int^{D+\epsilon}_D\sqrt{e^{Q/r^2}+C}~dr.
\end{eqnarray}
This equation is not integrable so one can not find its analytic
solution. We shell use the numerical technique and observe the
effect of electromagnetic field on proper length of the shell. Figure \eqref{b2} describes the direct link of the length of the shell with its thickness. It is also seen from Fig.\eqref{b2} that as we increase the charge within fluid, then the length of the shell will be decreased. This indicates that electromagnetic field is trying to lessen down the total length of the shell of a gravastar. Thus, uncharged gravastars have relatively longer length of the shell.

\subsection{Energy Content}

\begin{center}\begin{figure}\centering
\epsfig{file=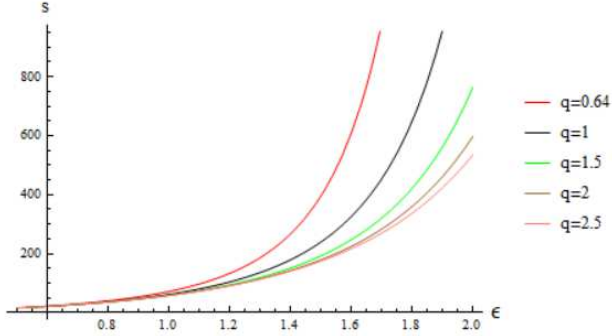,width=0.5\linewidth}\caption{\small{Plot
of $\varepsilon$(km) verses $\epsilon$ in the gravastars with
$\chi=1$.}}\label{b3}
\end{figure}\end{center}

In the interior region, we used a special combination of EoS given by $p=-\rho$ EoS which demonstrates the zone of negative energy that reflects the non-attractive nature of force at the interior geometry. The energy of the thin shell can be
given as
\begin{eqnarray}\label{43}
\varepsilon=\int^{D+\epsilon}_D4\pi\rho r^2~dr,
\end{eqnarray}
which turns out to be
\begin{eqnarray}\label{44}
\varepsilon= 4\pi \int^{D+\epsilon}_D r^2
Exp\left[-\frac{r^4+Qr^2-Q}{8Q}e^{\frac{Q}{r^2}}+\frac{r^2}{4}\right]~dr,
\end{eqnarray}
Again the integration of above expression is not possible so we can not find
the analytic solution. By using numerical method, one can observe the
effect of electromagnetic field on the energy content within the
shell. In Fig.\eqref{b3}, the energy $\varepsilon$ (km) inside the shell is plotted in relation to the shell thickness $\epsilon$ (km) in the presence of electromagnetic field. The plot \eqref{b3} indicates that less charged fluid has a relatively greater energy within the shell of the gravastars.

\subsection{Entropy within the Shell}

\begin{center}\begin{figure}\centering
\epsfig{file=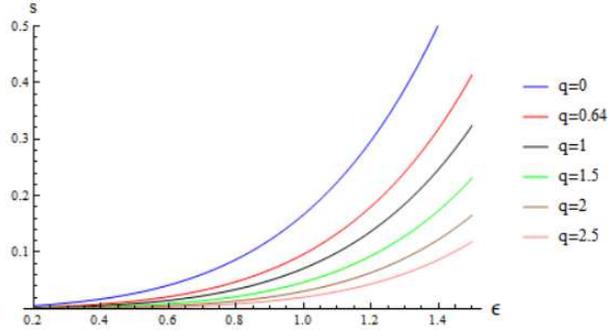,width=0.5\linewidth}\caption{\small{Plot
of $s$(km) verses $\epsilon$ in the gravastars with $\chi=1$,
$F=0.00001$, $C=0.00006$.}}\label{b4}
\end{figure}\end{center}

According to the theory of Mazur and
Mottola \cite{40}, the interior region of the gravastars has no
disorderness. Now we will find out the entropy by taking
$r_1=D$ and $r_2=D+\epsilon$ radii of interior and exterior regions, respectively.
According to the theory of Mazur and Mottola, entropy within the
shell can be calculated using following formula
\begin{eqnarray}\label{45}
s=\int^{D+\epsilon}_D4\pi r^2\Re(r)\sqrt{K}~dr,
\end{eqnarray}
where $\Re(r)$ is the entropy density which can be defined as
\begin{eqnarray}\label{46}
\Re(r)=\left(\frac{\alpha
K_B}{\hbar}\right)\left({\frac{P}{2\pi}}\right)^\frac{1}{2},~~~~~~~~~K_B=\hbar=1.
\end{eqnarray}
Using Eqs.\eqref{30} and \eqref{47} in Eq.\eqref{46}, the
entropy within the shell turns out to be
\begin{eqnarray}\label{47}
s=(8\pi
F)^\frac{1}{2}\alpha\int^{D+\epsilon}_Dr^5\sqrt{e^{Q/r^2}+C}~dr,
\end{eqnarray}
The figure has been plotted in order to analyze the disorderness against the shell thickness. Figure \eqref{b4} describes zero entropy
of the cylindrically symmetric charged gravastar-like relativistic structures at zero thickness of the fluid, which is one of the viable conditions
for the single condensate phase of the stellar bodies as described by \cite{40}.

\section{Conclusion}

This paper is devoted to understand the existence of gravastar under cylindrical symmetric metric in the realm of Maxwell-$f(R,T)$ gravity. For this purpose, we consider the cylindrical geometry and established the subsequent field equations and conservation laws in $f(R,T)$ gravity.
We studied the interior and exterior geometry of gravastar and examined the behavior of charge on the formulation of pressure and energy density of the gravastar. We connected the interior and exterior cylindrical region through some suitable matching conditions. Gravastar could be considered as a viable alternative to the black holes, which is the end state of gravitational collapse. Such an end state can be elaborated with the help of three regions containing specific EoS. First is an interior cylindrical region and the second is a middle thin shell and the third is exterior cylindrical region. We also discussed the viable features of gravastar and observed electromagnetic charge effects on it using cylindrically symmetric spacetime. We determined the mass function of the middle thin shell using interior and exterior geometries. The several salient results found in this work can be described in the following steps.\\
(i) $Density-Pressure~profile:$\\
 The comparison between the two important structural variables, i.e.,  pressure and density of the ultrarelativistic matter configuration within the intermediate thin shell are described in Fig.\eqref{b1}. This escribes a special variation in the profiles of matter variables for charge and radial coordinate. This also describes an abrupt change in the profiles of pressure and energy density for the low charged fields. \\
(ii) $Proper~length~of~thin~shell:$\\
The proper length of the thin shell is plotted with respect to its thickness described in Fig.\eqref{b2}, which indicates the progressively increasing profile. Furthermore, an uncharged medium occupies a high value of proper length and vice versa. From the graph \eqref{b2}, we can infer that if the charge inside the gravastar raises then the length of the thin shell falls and vice versa.\\
(iii) $Energy~content:$\\
We can see the effects of charge and thickness on the energy content within the thin shell through Fig.\eqref{b3}. It is analyzed that the shell energy has a direct proportion to its thickness. Moreover, the increase in the electric charge will directly increase shell thickness and increase of electric charge will decrease the amount of energy for the gravastars. \\
(iv) $Entropy:$\\
We have illustrated a diagram to investigate the function of electric charge and thickness on the entropy of the system. Figure \eqref{b4} indicates the increasing behavior of the entropy by decreasing the contribution of charge and vice versa. Furthermore, the entropy increases slowly with respect to shell thickness, thereby representing the highest value of thickness at hypersurface. Our analysis could be useful to understand the effects of electromagnetic field on cylindrically symmetric gravastar-like bodies in $f(R,T)$ gravity.

\vspace{0.5cm}

{\bf Acknowledgments}

\vspace{0.25cm}

This work was supported by National Research Project
for Universities (NRPU), Higher Education Commission, Islamabad
under the research project No. 8754/Punjab/NRPU/R\&D /HEC/2017.

\end{document}